\documentclass[amsmath,amssymb, aps,pre,preprint]{revtex4-1}

\usepackage{graphicx}
\usepackage{epstopdf}
\usepackage{dcolumn}
\usepackage{bm}
\usepackage{subfigure}

\usepackage[colorlinks]{hyperref}
\hypersetup{
citecolor=blue,
menucolor=blue,
linkcolor=blue,
urlcolor=blue,
}

\begin{document}


\title{Packing loops into annular cavities}

\author{T. A. Sobral\footnote{Present address: \textit{Department of Physics, Universidade Federal do Cear\'a, 60451-970 Fortaleza, Cear\'a, Brazil.}}}
\author{M. A. F. Gomes}
\email{mafg@ufpe.br}
\affiliation{Departamento de F\'isica, Universidade Federal de Pernambuco, 50670-901, Recife PE, Brazil}
\date{\today}


\begin{abstract}

The continuous packing of a flexible rod in two-dimensional cavities yields a countable set of interacting domains that resembles non-equilibrium cellular systems and belongs to a new class of light-weight material.
However, the link between the length of the rod and the number of domains requires investigation especially in the case of non-simply connected cavities, where the number of avoided regions emulates an effective topological temperature.
In the present article we report the results of an experiment of injection of a single flexible rod into annular cavities in order to find the total length needed to insert a given number of loops (domains of one vertex).
Using an exponential model to describe the experimental data we quite minutely analyze the initial conditions, the intermediary behavior, and the tight-packing limit.
This method allows the observation of a new fluctuation phenomenon associated with instabilities in the dynamic evolution of the packing process. Furthermore, the fractal dimension of the global pattern enters the discussion under a novel point of view.
A comparison with the classical problems of the random close packing of disks, and jammed disk packings is made.

\begin{description}
\item[Keywords] \keywords{} Folded structures, 2D systems, complex packing.
\end{description}

\end{abstract}

\maketitle


\section{Introduction}\label{secI}

The packing of a long flexible rod of diameter $\zeta$ into a finite quasi-two-dimensional domain of height $\gtrsim \zeta$ yields a network of loops which is a pattern of folds that belongs to a new class of light-weight cellular materials with variable degrees of rigidity and with great potential for application in technology~\cite{Donato02,Stoop08}. The general problem of a flexible rod confined in a two-dimensional cavity belongs to the context of elastic rods constrained to move on surfaces~\cite{Huynen15,Oshiri15}. The structures formed in these circumstances have a high surface/bulk ratio, and they extend the field of evolving non-equilibrium cellular systems composed of interacting domains separated by thin boundaries endowed with line energy~\cite{Stavans93}. New results on this subject may also be of interest in the study of shape memory materials using elastoplastic alloy wires~\cite{Gadot15} among others~\cite{Oppenheimer15}.

Since the rod bends when inserted into the cavity and divides the available area into geometric domains, we can perform a direct comparison between the continuous packing and the discrete number of cells. In this context, the present study deals with the length $L$ required to build a given number $n$ of loops (domains with one vertex). At the beginning of the injection, the number of loops is equal to the number of self-contact points along the rod, but with the evolution of the process the contacts extend along segments of non-zero measure. Different morphological conformations are observed for confined rods \cite{Stoop08,Vetter14}, and they depend on the plasticity and on the friction at the rod-cavity interface, as well as between different parts of the rod. For the tight-packing in a given area, it has been conjectured~\cite{Gomes10,Gomes13} that the topology of the cavity regulates an effective temperature of the system: the smaller the number of avoided regions, the greater the packed length, the higher the temperature. The following results stand out: (\textit{i}) the successful application of an exponential description~\cite{adda10,bayart11,Sobral15} also for annular cavities; (\textit{ii}) the observation of a new ``instability'' phenomenon; and (\textit{iii}) a new context in which the physical rigidity of the rod imposes an effective fractal dimension that is always smaller than 2 for the tight-packing conformations.

This paper is divided as follows: In Sec.~\ref{secII} the experiment of packing a single rod into annular cavities is detailed. An exponential model that fits the experimental data is discussed in some detail in Sec.~\ref{secIII}. Our results are reported in Sec.~\ref{secIV} from the point of view of the formation of each loop (Sec.~\ref{subsecIVa}) and from the point of view of tight-packing measurements (Sec.~\ref{subsecIVb}). The conclusions are summed up in Sec.~\ref{secV}.


\section{Experimental details}\label{secII}

\begin{figure}[!b]
\includegraphics[width=0.6\linewidth]{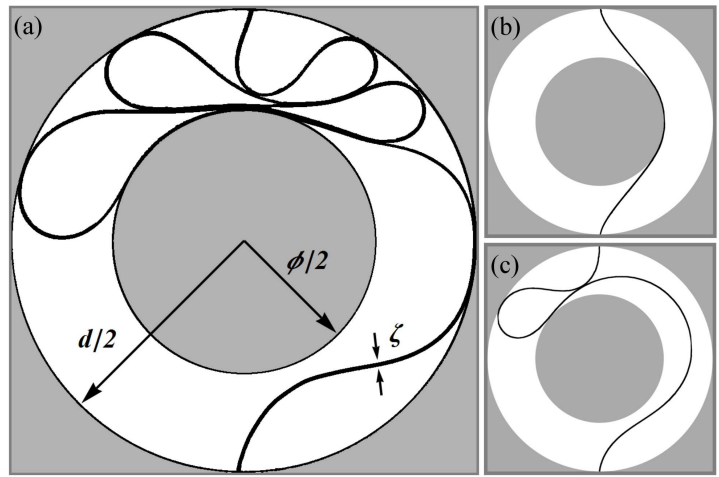}
\caption{\label{Fig1} (a) Generation of four loops from the packing of a flexible rod (diameter $\zeta$) inside an annular cavity of inner (external) diameter $\phi$ ($d$). (b) The initial conformation with $L_{IC}$ in length. (c) The formation of the first loop, whose length is $L_{IC} + \lambda_1^* = L_1 \equiv L_0 + \lambda_1$ [Eq.~(\ref{eq3})].}
\end{figure}

The cavity is composed of the superposition of two acrylic plates of 10~mm thickness. A circular groove of diameter $d=201$~mm allows us to accommodate only one layer of the flexible rod of diameter $\zeta=1.0$~mm. Two opposite parallel slits compose the injection channels. The rod is then packed into the cavity from an injection channel and can be recovered by the other. The cavity is the same as used in previous studies~\cite{Donato02}, but here we change its topology by adding a set of central aluminum disks (Fig.~\ref{Fig1}) with a selected interval of more than one decade in diameter $\phi = \{3, 15, 20, 27, 36, 48, 64, 85, 113, 150\}$~mm or more than three decades in the excluded area. 
The total area available to the rod is 
\begin{equation}
A(\phi) = \frac{\pi}{4} \left( d^2 - \phi^2 \right).
\label{eq1}
\end{equation}
For comparison, the experiment is also performed in a simply connected cavity (without the central disk). Inside mathematical expressions we mention simply connected cavity as ``scc''. It is important in this study that $A\textrm{(scc)}=A(0)$ in Eq.~(\ref{eq1}) because in this limit we can distinguish the contribution of the topology. The flexible rod is a typical nylon fishing line of diameter $\zeta = 1.0$~mm with a Young's modulus of about $Y = 2$~GPa.

The initial conformation is shown in Fig.~\ref{Fig1}(b).  At the outer ends of the rod a mark in made with a permanent marker. The distance between these marks indicates the length of the rod in this initial condition, $L_{IC}$. Once injected, the rod bends continuously changing its geometric pattern. A loop is a closed geometric teardrop shape composed of a vertex and a bulge. The first loop arises when the rod makes the first self-contact as shown in Fig.~\ref{Fig1}(c). During the injection more loops are created inside the cavity, with several sizes and positions. Our experiment consists of measuring the length needed to add a new loop to the total geometric pattern. When a new self-contact arises, the injection is stopped in order to mark the length on the rod with the marker. The injection speed is about $v = 1$~cm/s, but a pause is taken to dry the ink and to carefully remove the excess. All injections are performed in a dry regime.

Compared to the simply connected cavity the presence of a central obstacle changes the disposition of the loops to the periphery of the cavity. The region close to the injection channel presents a denser concentration of the rod than in the opposite region of the cavity, behind the obstacle, which becomes scarcely available to the rod even at small values of $\phi$. The size of the loops varies with the diameter of the central disk and therefore the total length needed to create $n$ loops depends on the area of the annulus.

We are interested in investigating the total length needed to create a given quantity of loops in a global geometric pattern. Assuming that the length needed to build $n$ loops is $L_n$, then the length $\lambda_{n+1}$ needed to create a new loop is
\begin{equation}
\lambda_{n+1} = L_{n+1}-L_{n}
\label{eq3}
\end{equation}
and the total length can be computed by summing up the distances between successive marks along the rod. The instrumental uncertainties are fixed for $\lambda_n$ and increase with $ \sqrt{n}$ for $L_n$ (although never more than $2\%$).

The injection ends when the system becomes jammed. In general the last mark on the rod does not indicate the tight-packing, because the system can become jammed after the formation of the last loop. The last mark corresponds to the total length needed to create $n_{max}=N$ loops in the cavity. After the injection, the rod is extracted through the opposite channel. The whole experiment consists of 10 identical realizations of packing of the flexible rod into an annular cavity. A total of 3907 loops are observed in 110 realizations.


\section{The exponential model}\label{secIII}

Earlier studies claimed that the relationship between the number of loops and the length of the rod follows a power law~\cite{Donato02,Stoop08}. However, the length of branches of the rod between successive self-contacts follows exponential distributions~\cite{adda10,bayart11}. For regular folding~\cite{deboeuf13}, the successive bending of a rod of length $\lambda_0$ in the middle yields $n_i$ pieces of length $\lambda_i$ between successive kinks, with $n_i = 2^{i}$ and $\lambda_i = \lambda_0/2^i$, which corresponds to a power-law scaling $\lambda_i = n^{-1}_i \lambda_0$. We have observed that, since each loop has an individual length ($n_i=1$), the number of loops must be identified as the iteration of the process, and therefore we propose an exponential scaling. Indeed, a recent study of the subject~\cite{Sobral15} introduces successfully an exponential law in order to describe the unpacking of a copper wire. In the present paper, an exhaustive experimental analysis substantiates the exponential description of the packing of a flexible rod and an examination of its limitations demonstrates instabilities not reported before. This approach also enables a different point of view about the fractal behavior usually found for tight-packing configurations.

\begin{figure*}[!ht]
\includegraphics[width=\linewidth]{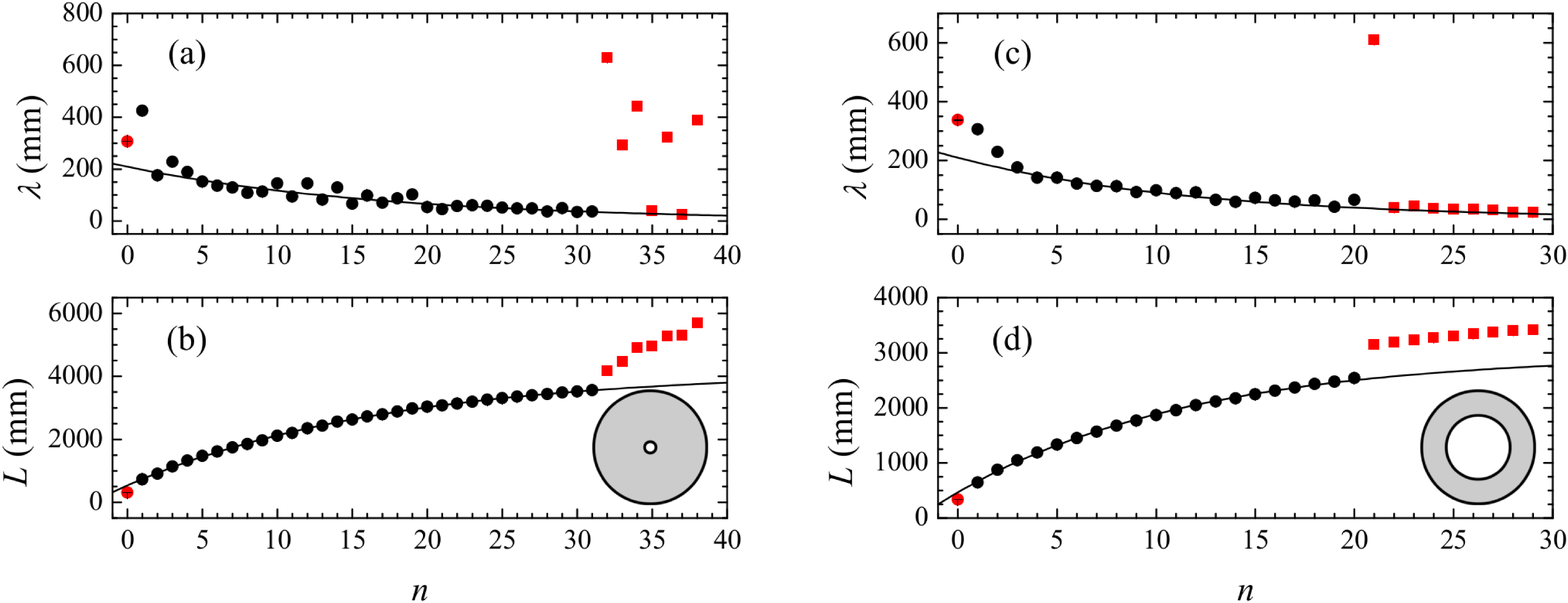}
\caption{\label{Fig2} (a, c) Length $\lambda_n$ between successive loops and (b, d) total length $L_n$ needed to create $n$ loops for annular cavities of (a, b) $\phi =20$~mm, and (c, d) $\phi=113$~mm. The first point ($n=0$) corresponds to the initial length [Eq.~(\ref{eq2})] and the squared marks ($n>N^*$) refer to lengths that deviate from the long course original behavior (see Sec.~\ref{secIII} for details). These are not considered in the exponential fitting (solid lines).}
\end{figure*}

Figures~\ref{Fig2}(a) and~\ref{Fig2}(c) show the length $\lambda_n$ and Figs.~\ref{Fig2}(b) and~\ref{Fig2}(d) show the total length $L_n$ as functions of the number of loops $n$ for generic injections with $\phi = 20$~mm~[Figs.~\ref{Fig2}(a) and~\ref{Fig2}(b)] and $\phi = 113$~mm~[Figs.~\ref{Fig2}(c) and~\ref{Fig2}(d)]. 
The first experimental point in Fig.~\ref{Fig2} corresponds to the injected length $L_{IC}$ which is associated with the initial condition of the rod in the annulus [Eq.~(\ref{eq2})]. However, the counting of loops must start from a given length $L_0$ which is initially unknown. As a consequence, one can not consider $L_{IC}$ for theoretical fits and one must find $L_0$ from extrapolation of all the data. 

The relationship between the cumulative length $L_n$ and the length between successive loops $\lambda_n$ can be written as
\begin{equation}
L_n = L_0 + \sum_{i=1}^{n} \lambda_i
\label{eq4}
\end{equation} 
and it is in agreement with Eq.~(\ref{eq3}) where $\lambda_1 = L_1-L_0$. Since $L_0$ is unknown, $\lambda_1$ is initially unknown and it is different from the experimental value, $\lambda_1^* = L_1-L_{IC}$. For annular cavities, generally $L_0>L_{IC}$ and the theoretical fit for $L_n$ passes above the first data point (see Fig.~\ref{Fig2}). As an immediate consequence, the value of $\lambda_1^*$ is higher than predicted by the theoretical line, i.e., $\lambda_1^*>\lambda_1$.

The data for $\lambda_n$ are very well described by an exponentially decreasing function of $n$, as expected from simple arguments. Consider the existence of a maximum length $L_M$ that can be injected inside the cavity. It is reasonable to suppose that the new loop is created by consuming a fraction of the available length $L_M-L_n$:
\begin{equation}
\lambda_{n+1} = \upsilon (L_M - L_n) , \hspace{20pt} (\upsilon<1).
\label{eq5}
\end{equation}
This equation can be iteratively used together with Eqs.~(\ref{eq3}) and~(\ref{eq4}) to obtain  
\begin{equation}
\lambda_{n} = \lambda_1 (1-\upsilon)^{n-1}.
\label{eq6}
\end{equation}
The total length $L_n$ [Eq.~(\ref{eq4})] needed to create $n$ loops is then a limited quantity. For the formation of infinite loops we have $L_\infty = L_M$. We can write
\begin{equation}
L_{n} - L_0= (L_\infty - L_0) \left( 1 - e^{-n/n_c} \right),
\label{eq7}
\end{equation}
where $n_c^{-1} = - \ln{(1-\upsilon)}$. Our model thus yields an equation with three physical parameters: (\textit{i}) $L_0$, a length of reference; (\textit{ii}) $L_\infty$, the saturation length; and (\textit{iii}) $n_c$, a characteristic number of loops. These parameters are obtained from fits. The solid lines in Figs.~\ref{Fig2}(b) and ~\ref{Fig2}(d) show the best fit of Eq.~(\ref{eq7}) over the experimental data. From the same fits the solid line in Fig.~\ref{Fig2}(a) and ~\ref{Fig2}(c) are obtained by applying Eq.~(\ref{eq3}).

\begin{figure}[!ht]
\includegraphics[width=0.6\linewidth]{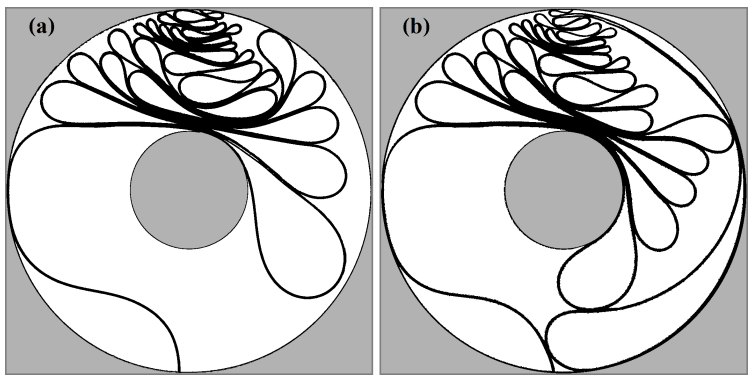}
\caption{\label{Fig3} Insertion of $n$ loops in an annular cavity of $\phi = 64$~mm and $d=201$~mm. (a) The loops decrease exponentially in size. (b) As the packing proceeds, the route of progressively smaller loops becomes unstable and new branches of larger loops can appear.}
\end{figure}

The exponential description works very well for the injection of the flexible rod into all cavities used in the experiment. The model is appropriate to describe the length of the rod which is adaptive to the whole structure within the cavity. In practice, the elastic interaction among branches of the rod is followed up by internal movements. Friction is responsible for intermittencies due to the stick-slip phenomenon. When the structure has enough loops and the insertion of a new one becomes uncertain due to the high rigidity, two situations are observed: (\textit{i}) the system is tightly packed, and it is impossible to inject the rod further; or (\textit{ii}) the structure slides and a new area of the cavity becomes accessible (Fig.~\ref{Fig3}). We denote the latter case ``instability''. Naturally, after one or more instabilities the system becomes jammed at some stage of the packing.

Returning to Fig.~\ref{Fig2}, we can now examine a consequence of these instabilities: the length $\lambda_{n+1}$ needed to insert the next loop is relatively large. As a consequence, there is a discontinuity in $L_n$. In our experiment this occurs rather unpredictably (36\% of realizations), and to the best of our knowledge, it has not been reported before in the literature on confined wires. The dry injection and the roughness of the cavities are pointed out as possible promoters of those instabilities. In cases of discontinuities, the theoretical fit is performed until a given number $N^*\leq N$, immediately before the instability. The values of $N^*$ vary between $N^* =(38.3 \pm 0.7)$ for larger cavities (with $\phi=15$~mm) and $N^*=19.3 \pm 0.5$ for small cavities (with $\phi = 150$~mm). Our experiment was projected to precisely evaluate the exponential growth, and discontinuities were found as unexpected (unpredictable) outcomes. For this reason, although we can carry out a qualitative discussion, we do not have enough data for a suitable investigation focused on those instabilities.

While the total length of the rod plays the central physical role in determining the mass or the global rigidity of the system,  the number of loops plays the role of time in the dynamic, in the sense that all physical quantities evolves when new loops are created. This model treats the packing of a rod as a discrete dynamical system.


\section{Results and discussions}\label{secIV}

We discuss our results in two parts. The formation of loops is modeled by an exponential growth given by Eq.~(\ref{eq7}) for all 110 experimental realizations until a given number of loops $N^*$. In Sec.~\ref{subsecIVa} we discuss how the parameters of this exponential fitting change when the inner diameter $\phi$ of the annulus varies. Since the injection becomes progressively harder the system jams at some stage of the packing. In Sec.~\ref{subsecIVb}, we discuss the dependence of the number $N^*$ with the total area of the cavity. This allows us to understand why all experiments on packing in two-dimensional cavities present a complex structure whose fractal dimension $D$ is slightly smaller than 2.


\subsection{Formation of loops}\label{subsecIVa}

\begin{figure}[!ht]
\includegraphics[width=0.7\linewidth]{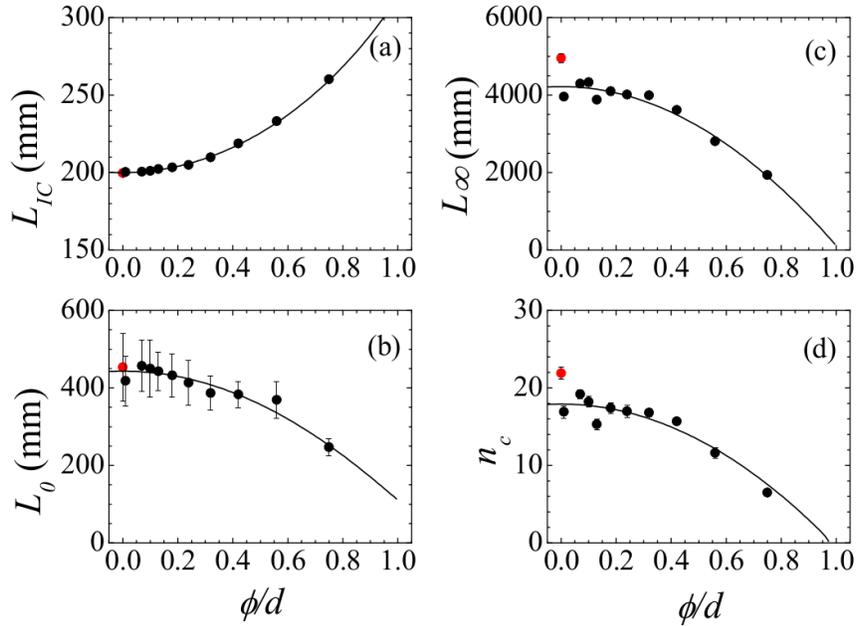}
\caption{\label{Fig4} (a) The initial length $L_{IC}$ increases with the inner diameter $\phi$ as expected from inspection of Fig.~\ref{Fig1}(b). However, all fit parameters in Eq.~(\ref{eq7}) -- (b) $L_0$, (c) $L_{\infty}$, and (d) $n_c$ -- decrease with $\phi$. The first data point corresponds to the simple connected cavity. The solid lines are curves obtained from the theoretical description: (a) Eq.~(\ref{eq2}), (b) Eq.~(\ref{eq9}), (c) Eq.~(\ref{eqA}), and (d) Eq.~(\ref{eq10}).}
\end{figure}

We described the initial length $L_{IC}$ corresponding to Fig.~\ref{Fig1}(b) by two components: a straight one, which links the channels to the (tangent) inner disk; and a curved one, which follows the circular arc of the central disk. The resulting expression,
\begin{equation}
L_{IC} = d \left\{ \sqrt{1-(\phi/d)^2} + (\phi/d) \arcsin{(\phi/d)} \right\},
\label{eq2}
\end{equation}
describes very well the length $L_{IC}$ for any diameter $\phi$, and also for the simply connected cavity $L_{IC}\textrm{(scc)} = d$, as shown in Fig.~\ref{Fig4}(a) by the solid line. Please observe that Eq.~(\ref{eq2}) is not obtained through a fit from experimental data but, instead, from a simple geometrical reasoning.

The parameters $L_0$, $L_\infty$, and $n_c$ in Eq.~(\ref{eq7}) are obtained from a single fit over $N^*$ experimental points and then averaged among 10 realizations. The outcome is shown in Fig.~\ref{Fig4}(b),~\ref{Fig4}(c) and~\ref{Fig4}(d), where their numeric values are shown. The continuous lines are not fits but obtained from more fundamental quantities which depend only on features of the rod and cavity. In the next paragraphs we focus on these quantities.

The network of loops starts with the creation of the biggest loop. In simply connected (circular) cavities it is expected that the length needed to create the first loop, $\lambda_1$, is proportional to the size of the cavity (the pattern is simply rescaled), so $\lambda_1 \sim d$~\cite{Donato02}. Here we discuss the impact of a single obstacle in the center of the cavity. 

\begin{figure}[!t]
\includegraphics[width=0.6\linewidth]{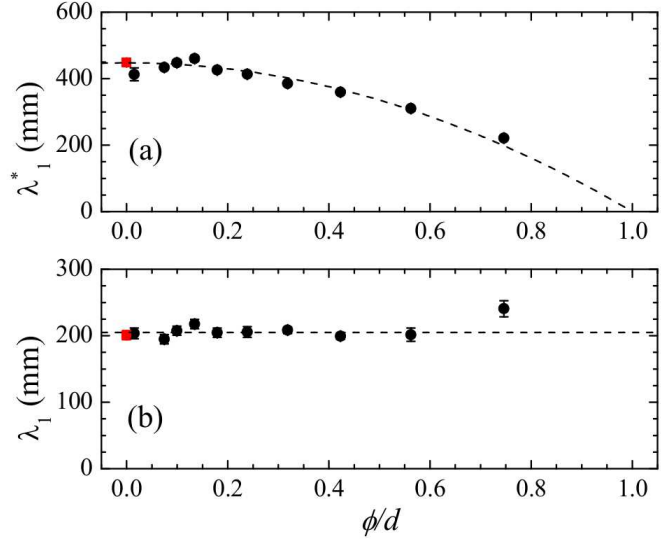}
\caption{\label{Fig5} The length needed to build the first loop. (a) The directly measured value $\lambda^*_1$ is proportional to the available area of the cavity, $\zeta \lambda^*_1 = 0.0141 A(\phi)$. (b) The model yields a fixed length, $\lambda_1 = 205$~mm.}
\end{figure}

Figure~\ref{Fig5}(a) shows the directly measured length $\lambda^*_1$ and Fig.~\ref{Fig5}(b) shows the proper length $\lambda_1$ as functions of the inner diameter of the annulus. The directly measured length $\lambda^*_1$ is found to be proportional to the available area of the cavity, $\zeta \lambda^*_1 = (0.0141\pm0.0004) A(\phi)$, as shown by the dashed line in Fig.~\ref{Fig5}(a). The first point corresponds to the first loop in a simply connected cavity whose value is the same, $\zeta \lambda^*_{1}\textrm{(scc)} = (0.0141\pm0.0003) A(0)$. The value for $\lambda_1$ is computed from the parameters of the exponential fitting
\begin{equation}
\lambda_1 = (L_\infty-L_0)\left(1 - e^{-1/n_c} \right)
\label{eq8}
\end{equation}
and it is found to be fixed, $\lambda_1 = (205\pm7)$~mm, as shown by the dashed line in Fig.~\ref{Fig5}(b). The first point corresponds to the first loop in a simply connected cavity whose value is the same within the error bars, $\lambda_1\textrm{(scc)} = (201\pm7)$~mm. This suggests that, although $\lambda_1$ depends on the external diameter $d$, it is unresponsive to the inner diameter $\phi$ of the annulus. All together, these results allow us to write
\begin{equation}
L_0(\phi) = \lambda^*_1(\phi) -\lambda_1(\phi) +L_{IC}(\phi)
\label{eq9}
\end{equation}
as a function of the inner diameter $\phi$. Equation~(\ref{eq9}) describes implicitly the dependence of the fitting parameter $L_0(\phi)$ and produces the solid line shown in Fig.~\ref{Fig4}(b). We can write $L_0\textrm{(scc)} = L_0(0)$ for the simply connected cavity since similar relations are valid for each term on the right-hand side of Eq.~(\ref{eq9}). Therefore, this parameter does not change between the two topologies.

\begin{figure}[!t]
\includegraphics[width=0.6\linewidth]{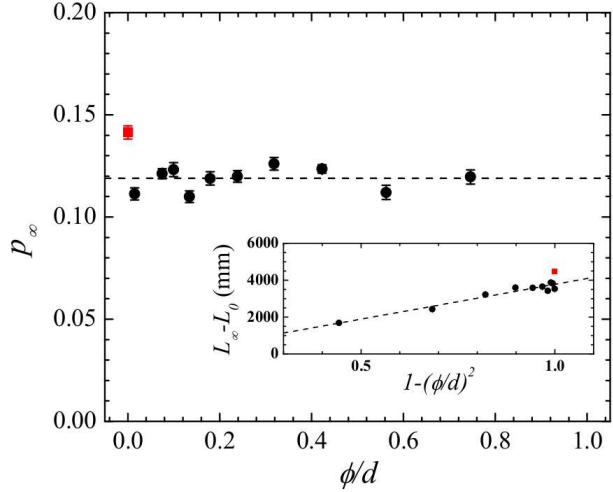}
\caption{\label{Fig6} The packing fraction is approximately fixed for all annular cavities, $p_\infty = (0.119 \pm 0.002)$. The first point is relative to the simply connected cavity, which has a bigger packing fraction $p_\infty\textrm{(scc)} = (0.142 \pm 0.003)$.  Inset: The interval of existence of loops $L_\infty - L_0$ is found to be proportional to the available area of the annulus.}
\end{figure}

The exponential model predicts a saturation length $L_M=L_\infty$ which can be obtained by fitting the experimental data for each annular cavity of inner diameter $\phi$. The pattern of infinite loops inside the cavity is regarded as a mathematical fractal, which is impossible to reach by physical constraints imposed by the plasticity and thickness of the rod. The range of the exponential model in length, $L_\infty-L_0$, is found to be proportional to the available area of the annulus (see inset in Fig.~\ref{Fig6}). From the proportionality constant we define the packing fraction $p_\infty$, which is approximately fixed for all annular cavities (Fig.~\ref{Fig6}). The packing fraction $p_\infty$ corresponds to the fraction of available area covered by the rod in the ideal configuration of infinity loops. This conformation characterizes the thermodynamic limit of this system. The saturation length
\begin{equation}
L_\infty(\phi) = L_0(\phi) + p_\infty \left( \frac{A(\phi)}{\zeta} \right)
\label{eqA}
\end{equation}
varies linearly with the available area in the limit of larger cavities ($L_\infty \gg L_0$). Equation~(\ref{eqA}) describes the dependence of the fitting parameter $L_\infty(\phi)$ and produces the solid line shown in Fig.~\ref{Fig4}(c). The dashed line in Fig~\ref{Fig6} illustrates the mean value $p_\infty = (0.119 \pm 0.002)$ found for annular cavities with inner diameter $\phi$.

In the examination of the ``stacking'' of loops of decreasing size on top of each other, it is quite interesting that the same area fraction of the cavity is filled, no matter how large the obstacle in the middle. Using polar coordinates whose origin coincides with the center of the annulus, we observe that a fixed packing fraction $p_\infty$ can be well explained by treating the rod (with infinite loops) as covering a fixed angular sector. Of course, this picture is inappropriate for a circular cavity without obstacles. The first point in Fig.~\ref{Fig6} represents the data for a simply connected cavity whose packing fraction is $p_\infty\textrm{(scc)} = (0.142 \pm 0.003)$. This result, $p_\infty\textrm{(scc)} \neq p_\infty(\phi \rightarrow 0)$, quantifies the impact of a non-simply connected topology on the total number of states accessible for the rod. It was conjectured by Gomes \textit{et al.}~\cite{Gomes10} that the packing of the rod can be described by a non-equilibrium thermodynamics where the number of obstacles regulates the effective temperature $T$ of the system. In this sense, the data from annular cavities are dispersed over an isotherm, and any change in the total length of the rod is a contribution uniquely of the available area.

The results illustrated in Figs.~\ref{Fig5} and~\ref{Fig6} allow us to write the dependence of the exponential parameters on the inner diameter $\phi$. The characteristic number of loops as defined in Eq.~(\ref{eq7}) is a fitting parameter whose dependence on $\phi$ is very well described by
\begin{equation}
n_c^{-1} = - \ln{\left\{ 1 - \frac{\zeta \lambda_1}{p_\infty A(\phi)} \right\}},
\label{eq10}
\end{equation}
as shown by the solid line in Fig.~\ref{Fig4}(d). The characteristic number of loops depends on the physical properties of the material through $\lambda_1$ and $p_\infty$. Together with the ideas previously discussed, where $\lambda_1 \sim d$ and $p_\infty \sim \textrm{fixed}$, Eq.~(\ref{eq10}) has two interesting physical limits. For larger cavities, the thermodynamic limit gives us that $n_c \rightarrow p_\infty / p_1$, where $p_1$ is defined as the packing fraction associated with the first loop, $p_1 \equiv \frac{\zeta \lambda_1}{A(\phi)}$. In the opposite limit, if the cavity is too narrow such that $p_1>p_\infty$ then the exponential behavior falls off and loops must no longer be observed! Indeed, this special situation was found in a recent study using rectangular cavities~\cite{Sobral15b}. For the set of annular cavities discussed here, we estimate that the central disk must have approximately $97\%$ of the diameter $d$ of the cavity in order to inhibit the formation of loops.

The exponential model described in this article is in agreement with previous studies of tight-packing in crumpled systems and regular folding, and it yields some predictions. The equality
\begin{equation}
\exp \left[-\frac{n}{n_c} \right] = 1 - \left( \frac{L_n-L_0}{L_\infty-L_0} \right) = \frac{\lambda_n}{\lambda_0},
\label{eqB}
\end{equation}
where $\lambda_0=\lambda_1 \, e^\frac{1}{n_c}$, has a connecting role between these approaches. In the regular folding case, the force scales with the compaction as $F \sim (\lambda_0 / \lambda_n)^{\beta_1}$~\cite{deboeuf13}, because the force increases in order to bend progressively smaller pieces of the rod. On the other hand, in the jamming approach the velocity of the injection decreases as the density of the rod increases close to the injection channel. The physical effect is that the force scales with the total length inside the cavity as $F \sim (1 - l / l_c)^{-\beta_2}$~\cite{Stoop08}, where we can identify $l\rightarrow L_n-L_0$ and $l_c\rightarrow L_\infty-L_0$. Following Eq.~(\ref{eqB}) both approaches are suitable for describing this system, and we find $\beta_1=\beta_2\equiv \beta>0$. We can also estimate the dependence of $n_c$ on the bending stiffness $\mu$ of the rod as follows. The force associated with the first loop must be proportional to the bending stiffness; then $F_1 \sim (\lambda_0 / \lambda_1)^{\beta} = e^{\frac{\beta}{n_c}} \sim \mu$, that is, $n_c^{-1}$ must increase linearly with $\ln{\mu}$. This prospect means that $n_c$ is a very slowly decreasing function of $\mu$. 


Our non-completely ordered packing of loops in two dimensions has a complementary relation with the amorphous disk packings originally called ``random close packed'', whose densities are $\rho\textrm{(rcp)}\approx0.83$($0.84$) for the monodisperse (bidisperse) case \cite{Berryman83}. More recently, amorphous collectively jammed disk packings dominated by large triangular grains seem to have significantly higher densities, close to $\rho\textrm{(jdp)}\approx0.88$~\cite{Donev04}. Coincidentally, the packing fractions $p_\infty(\phi)=0.119\pm0.002$, averaged over 10 different genus 1 cavities, and $p_\infty\textrm{(scc)}=0.142\pm0.003$, for genus 0 cavities shown in Fig.~\ref{Fig6}, are closely complementary (that is, $p_\infty \approx 1-\rho$) to the amorphous arrangements of circles studied. 

The discontinuity between the packing densities in simply connected cavities and in cavities with a central excluded domain, shown in Fig.~\ref{Fig6}, where $p_\infty\textrm{(scc)}- p_\infty(\phi) = 0.023 \pm 0.005$, is interpreted as a difference in the entropy between the configurations, once the entropy scales linearly with the length of the injected rod and, consequently, with the packing density \cite{Gomes10}. The inclusion of an excluded region, irrespective of its size, lowers the entropy of the packing. In this view, that discontinuity is connected with a first-order phase transition, as the entropy is the first derivative of the associated free energy: it is a topological transition; the existence of an exclusion region, even of size $\phi \rightarrow 0$, leads to a configuration of the rod with a substantially different packing density compared with that obtained with the simply connected cavity.

This section concerns the lengths associated with $n$ loops in the intermediary (before the instabilities) stage of the packing. The quantities $\lambda_1$ and $p_\infty$ were obtained from fit parameters [Eq.~(\ref{eq7})], while $\lambda_1^*$ and $L_{IC}$ are directly measured along the rod. It was observed that these geometrical quantities are dependent only on properties of the rod and cavity (available before the experiment) and, therefore, are more fundamental quantities. Conversely, the fit parameters in Eq.~(\ref{eq7}) can be estimated from these quantities using Eqs.~(\ref{eq9} -- \ref{eq10}). This allows the estimation of the length associated with $n$ loops for any annular cavity of inner diameter $\phi$. Our results stress that the topology of the cavity only affects $p_\infty$, that is, the porosity of the global pattern.


\subsection{Jamming and instabilities}\label{subsecIVb}

As an implication of a fixed packing fraction $p_\infty$, the fractal dimension of the global pattern in the infinite limit might be $D_\infty = 2$. This is admissible from the point of view of a rod distributed in a fixed angular sector of the annulus, which has a two-dimensional shape. However, a previous study pointed out a slightly smaller fractal dimension, $D=1.9\pm0.1$, for the tight-packing of copper wires into circular cavities~\cite{Donato02}. Another investigation found $D=1.8\pm0.1$ for a more plastic wire of the alloy Sn$_{0.60}$Pb$_{0.40}$, but presented theoretical reasoning predicting that this exponent should be 2~\cite{maycon08}, linking this system with long molecular chains densely packaged as DNA in chromosomes, or in virus capsids.  In this subsection we examine why these studies found $D\lessapprox2$ for the tight-packing of wires in the context of the exponential model described above. In short, we point out that the physical limit for the force implies that the system gets jammed with a finite number of loops $N^*$, and then the total length under this condition $L^*$ does not scale linearly with the available area. There is also a spatial limit. We can see the reason or even estimate the jamming of the system through the radii of curvature for the smallest loops, immediately next to the injection channel, which are of the order of the gauge $\zeta$ of the rod.

The fits analyzed in Sec.~\ref{subsecIVa} are valid until a given number of loops $N^*$, at which either the system gets jammed or an instability occurs. The determination of $N^*$ is easily visually performed from graphs as in Fig.~\ref{Fig2}. A more objective criterion can be elaborated, for instance, calculating the relative growth of the loops, $r = (\lambda_{n+1}-\lambda_n) / \lambda_n = -(1-e^{-1/n_c})$, a negative quantity supposed to be independent of $n$. In practice, the relative growth is a small fluctuating quantity. This quantity is suitable to detect instabilities because in such cases the length $\lambda_{n+1}$ is large compared to $\lambda_n$. We choose a threshold value, $r_c=2.3$, which determines the occurrence of instabilities: if $r > r_c$, then the jump is detected and $N^*$ is recorded. Throughout the whole experiment the choice $r_c=2.3$ agrees $98\%$ with the visual criteria. Smaller values, $r_c<2.3$, would indicate statistical fluctuations as discontinuities and larger values, $r_c>2.3$, would not detect some of the conspicuous instabilities.

\begin{figure}[!hb]
\includegraphics[width=0.6\linewidth]{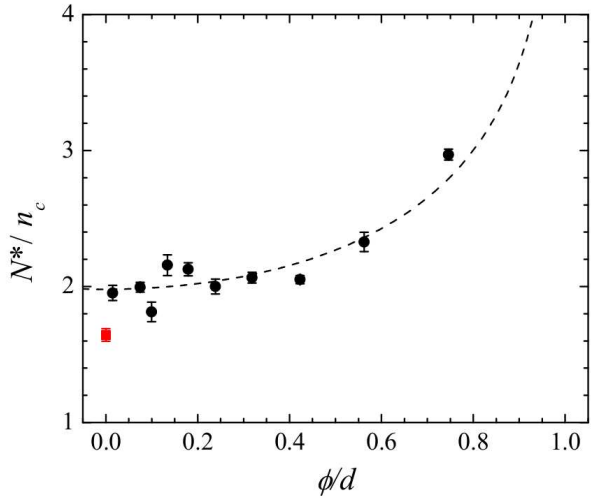}
\caption{\label{Fig7} The ratio $N^*/n_c$ illustrates the range of validity of the exponential fitting in the experimental data. In realizations on annular cavities, the exponential fit becomes better when the annulus is narrow (larger $\phi/d$).}
\end{figure}

Each realization of the experiment yields a particular geometric pattern and a total number of loops $N$, which vary considerably among different reproductions. On the other hand, the numbers of loops $N^*$ are close to each other such that we can compute an average value in order to illustrate all similar realizations with a given diameter $\phi$. We found $N^*$ as a decreasing function of $\phi$ similar to the characteristic number of loops $n_c$ [Eq.~(\ref{eq10})]. Figure~\ref{Fig7} shows the ratio $N^* / n_c$ as an increasing function of the diameter $\phi/d$, which reveals a relatively better fit for a narrower annulus.

\begin{figure}[!t]
\includegraphics[width=0.6\linewidth]{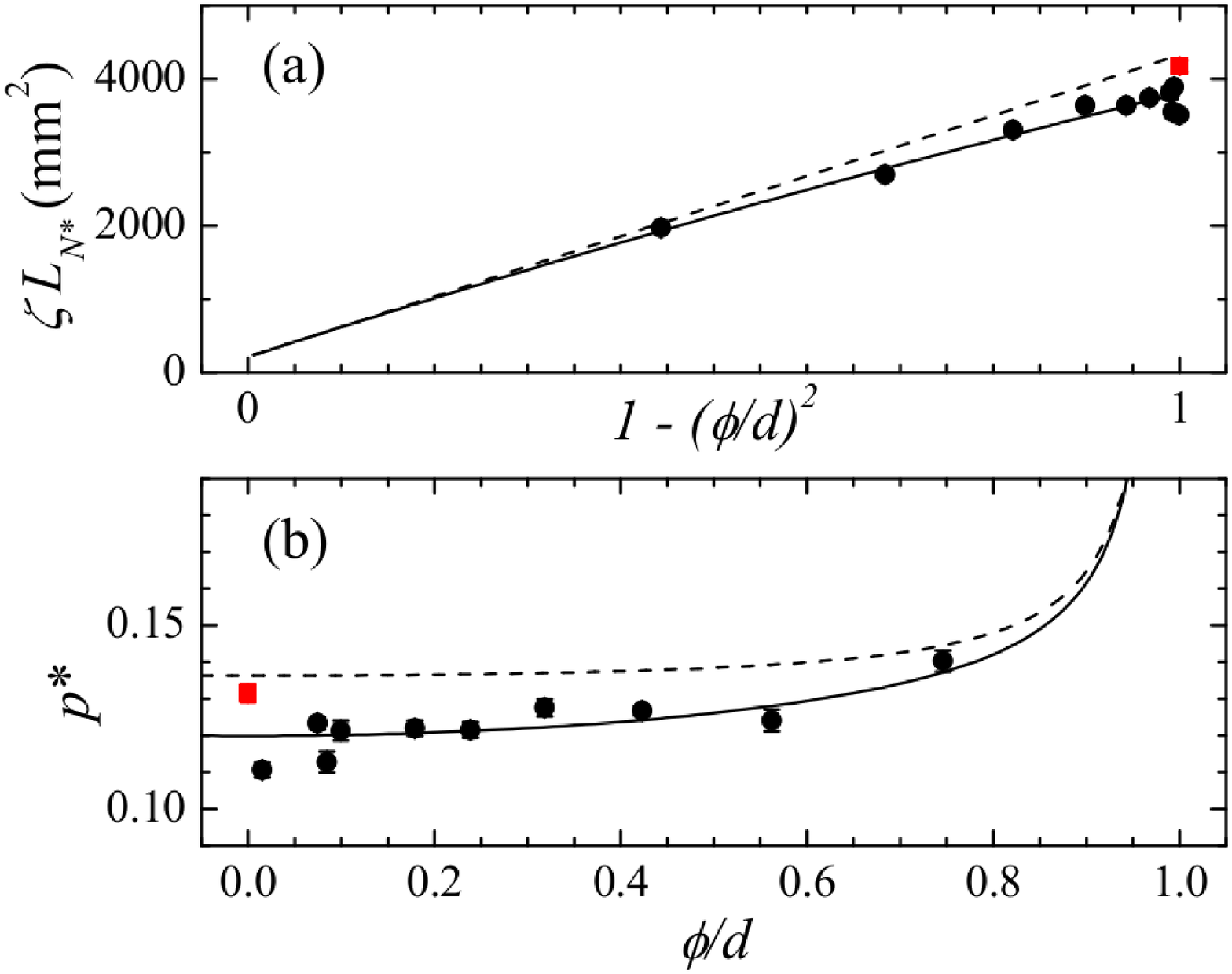}
\caption{\label{Fig8} tight-packing measures. (a) Area $\zeta L_{N^*}$ covered by the rod inserted in the cavity and (b) corresponding packing fraction $p^*$. The dashed line indicates the limit of infinite loops and the solid line shows the theoretical value given by Eq.~(\ref{eq12}).}
\end{figure}

The total length inserted in order to generate $N^*$ loops in the global packing pattern is $L_{N^*}$, given by Eq.~(\ref{eq7}) with $n=N^*$. Since $N^*/n_c$ is not fixed among the cavities, the total length $L_{N^*}$ is not simply proportional to the available area of the cavity. From the experimental data we found that the ratio $N^*/n_c$ linearly decreases with the logarithm of the available area, that is,
\begin{equation}
\frac{N^*}{n_c} = C_1- \ln{\left( 1-\left( \frac{\phi}{d}\right)^2 \right)},
\label{eq11}
\end{equation}
where $C_1 = 1.98\pm0.06$ is found by the best fit (see dashed line in Fig.~\ref{Fig7}). As a direct consequence, we can write
\begin{equation}
\zeta L_{N^*} = \zeta L_0(\phi) + p_ \infty A(\phi) - \left( \frac{p_\infty e^{-C_1}}{A(0)}\right) A(\phi)^2,
\label{eq12}
\end{equation}
where $p_\infty$, $C_1$ and $A(0)=\frac{\pi d^2}{4}$ do not depend on $\phi$. The plot in Fig.~\ref{Fig8}(a) shows the experimental data and the curve described by Eq.~(\ref{eq12}) (solid line). In the packing of infinite loops, the last term in Eq.~(\ref{eq12}) vanishes [see Eq.~(\ref{eq7})] and the length $L_{N^*}$ is essentially proportional to the available area (see the dashed line in Fig.~\ref{Fig8}).

We observe that Eq.~(\ref{eq12}) can be understood as the relation between the total mass of the flexible rod, $M \sim \zeta L$, and the size of the cavity, $R \sim \sqrt{A(\phi)}$, in the tight-packing limit. From the mass-size relationship, we can obtain the fractal dimension $D$ of the global pattern as $M \sim R^D$. Equation~(\ref{eq12}) suggests that the fractal dimension of a pattern of infinite loops is $D_\infty=2$. However, the physical limit of rigidity in the tight-packing imposes a deviation from the configuration of infinite loops and the tight-packing of a flexible rod in two-dimensional cavities gives an apparent fractal dimension $D \lessapprox 2$ as reported in the literature~\cite{Donato02,Gomes10}. The plot in Fig.~\ref{Fig8}(b) shows the packing fraction $p^* = \zeta L_{N^*} / A(\phi)$ as a function of the inner diameter of the annulus. From tight-packing measures, our result shows that the packing fraction is an increasing function when the initial length $L_0$ is ignored. This is the opposite effect of increasing the number of obstacles, which decreases the total packing fraction of the rod~\cite{Gomes10,Gomes13}. A divergence for $d-\phi  = 2\zeta$ is expected because an annulus of this width yields an initial packing fraction of $p_0 \approx 0.5$.

Although we can not study the instabilities in greater detail, we report some results of interest. The number of experimental realizations with instabilities does not shown any tendency when the inner diameter $\phi$ changes. Among all the realizations for a given $\phi$, there was at least one with instabilities, and on the other hand, there were at least three cases with no instabilities at all. The length of the rod inserted from the instability to the last loop is $\Delta L = L_N-L_{N^*}$ and it accumulates all discontinuities in the range $N^*\leq n \leq N$. On average, $\Delta L$ corresponds to an increment  $\Delta p = (0.062\pm0.005)$ in the packing fraction. However, the unpredictable behavior after the instabilities allows few realizations where $\Delta L$ can sum to about $100\%$ of the length inserted in the exponential domain $0<n \leq N^*$. We point out that the exponential behavior is suitable for describing the formation of loops with the injection of flexible elastic rods into two-dimensional cavities. The global patterns have a clear size gradient characteristic of what has been called \textit{classical morphology}~\cite{Stoop08}. For less elastic filaments, one expects more instabilities to occur. We stress that further studies of those discontinuities will lead to better descriptions of the formation of loops in patterns of \textit{plastic morphology}.


\section{Conclusions}\label{secV}

In this paper we report an exhaustive experimental study and analysis of the formation of loops in the packing of a flexible rod into annular cavities. The packing is treated as a discrete dynamical process that obeys an exponential model valid in simply connected cavities as well as in annular cavities. The length of formation of the first loop $\lambda_1$ and the length of the rod needed to build infinite loops $L_\infty$ define the exponential parameters, given a reference length $L_0$ where the loops starts counting. The model points out that there are discontinuities in the total length for a number of loops $n$ larger than the threshold $N^*$ in $36\%$ of realizations. The tight-packing is also analyzed and we show that the physical limit for the rigidity of the system imposes an apparent fractal dimension that is always smaller than two as previously found in different experimental studies. Our results in annular cavities are in agreement with previous claims which include changes in entropy due to the topology of the cavity and constant packing fraction for the effective isotherm curve of this system~\cite{Gomes10}. Approximate complementary relationships of the packing problems discussed here with the classical problems of the random close packing of disks, and the jammed disk packing are pointed out. We urge that additional study of the system examined in this article is needed in order to clarify the influence of plasticity and elasticity in the phenomenon and, in particular, to further the examination of the origin of the discontinuities reported here, and the possible connection of the appearance of instabilities with traditional studies of dynamical systems.


\acknowledgements

The authors acknowledge important suggestions from two anonymous referees. T. A. S. thanks Professor Jos\'e A. Soares Jr., Universidade Federal do Cear\'a, Fortaleza, Brazil, for his hospitality.
This work was supported by Grant No. 302251/2010-3 of CNPq (Conselho Nacional de Desenvolvimento Cient\'ifico e Tecnol\'ogico) and by Grant No. MCT/CNPq/FACEPE, APQ-1330-1.05/10 of PRONEX (Programa de N\'ucleos de Excel\^encia), all Brazilian agencies.


%

\end{document}